\documentclass[preprint,aps ,nofootinbib]{revtex4}
\usepackage{graphicx}
\usepackage{amsmath}
\usepackage{amsfonts}
\usepackage{amssymb}
\usepackage{color}%
\usepackage{dcolumn}
\providecommand{\U}[1]{\protect\rule{.1in}{.1in}}

\newcommand{\f}{\begin{equation}}
\newcommand{\ff}{\end{equation}}
\newcommand{\fa}{\begin{eqnarray}}
\newcommand{\ffa}{\end{eqnarray}}

\begin{document}
\title{The St\"{u}ckelberg Holographic Superconductors with Weyl corrections}
\author{Da-Zhu Ma $^{1}$}
\email{mdzhbmy@126.com}
\author{Yue Cao $^{2}$}
\email{yuecao860408@gmail.com}
\author{Jian-Pin Wu $^{2}$}
\email{jianpinwu@mail.bnu.edu.cn}
\affiliation{$^1$School of science, Hubei University for Nationalities, Enshi 445000, China
\\$^2$Department of Physics, Beijing Normal University, Beijing 100875, China}

\begin{abstract}

In this letter we construct the St\"{u}ckelberg holographic superconductor with Weyl corrections.
Under such corrections,
the Weyl coupling parameter $\gamma$ plays an important role in the
order of phase transitions and the critical exponents of second order phase transitions. So do the model parameters $c_{\alpha}$, $\alpha$ and $c_{4}$.
Moreover, we show that the Weyl coupling parameter $\gamma$ and
the model parameters $c_{\alpha}$, $\alpha$, $c_{4}$ which together
control the size and strength of the conductivity coherence peak
and the ratio of gap frequency over critical temperature $\omega_{g}/T_{c}$.

\end{abstract} \maketitle

\section {Introduction}

As is well known, there exist lots of questions in strongly coupled field theories. Facing these problems, many physicists resorted to the AdS/CFT correspondence\cite{Maldacena1997,Gubser1998,Witten1998,MaldacenaReview}, a equivalence between a $d$-dimensional
strongly coupled gauge theory on the boundary and a $(d + 1)$-dimensional weakly
coupled dual gravitational description in the bulk, to deal with some unsolved problems. Nowadays, as an application of the correspondence, the AdS/Condensed Matter duality (AdS/CMT) is also being developed. One remarkable achievement of AdS/ CMT is the establishment of the holographic superconductors \cite{3H}. It originated from the discovery of Abelian Higgs mechanism,
the $U(1)$ gauge symmetry is spontaneously broken
due to the existence of a black hole \cite{OriginalDiscovery1,OriginalDiscovery2}.
For a general review of the subject, please see \cite{HorowitzReview,HartnollReview1,HartnollReview2,HerzogReview,McGreevyReview}.

Of course, this model also displays some interesting properties. For example, there is a second order phase transition which is the standard behavior shown by Landau-Ginzburg theory.
Another important feature is that the ratio of gap frequency over critical temperature $\omega_{g}/T_{c}\approx 8$,
which is much larger than the value of $3.5$ given by the weakly coupled BCS theory,
but it is quite similar to that in the high-$T_{c}$ superconductors experiment \cite{GapNature}.

Recently, Franco $et$ $al.$ has proposed a phenomenological model,
in which the spontaneous breaking of a global $U(1)$ symmetry occurs
via the St\"{u}ckelberg mechanism\cite{stuckelberg1,stuckelberg2,stuckelberg3}.
They regard this model as the St\"{u}ckelberg holographic superconductor which has lots of features.
One main characteristic of this phenomenological model is that
it can provide a description of a large group of phase transitions,
including the first order phase transition and
the second order phase transition with both mean and non-mean field behavior.
Many extending works on the St\"{u}ckelberg holographic superconductor have been explored,
such as the St\"{u}ckelberg holographic superconductor
with Gauss-Bonnet corrections, the St\"{u}ckelberg holographic superconductor with backreactions \cite{stuckelberg4,stuckelberg5},
and the St\"{u}ckelberg holographic superconductors in constant external magnetic field \cite{stuckelberg6}.
In Ref.\cite{stuckelberg4}, they find that different values of GB parameter and model parameters
can determine both the order of phase transitions and the critical exponents of second order phase transitions.
What's more, the size and strength of the conductivity coherence peak can be controlled.

In this paper, we pay attention to the effects of the
Weyl corrections, another higher derivative corrections on the St\"{u}ckelberg holographic superconductor.
The Weyl corrections involve a coupling between the Maxwell field and the bulk Weyl tensor.
In Ref. \cite{WeylM}, the authors calculated the conductivity and charge diffusion with Weyl corrections, and
showed that corrections broke the universal relation with the $U(1)$ central charge observed at leading order.
In addition, the holographic superconductors with Weyl corrections are also explored in Ref.\cite{JPWu}.
They found that for large Weyl parameter $\gamma$,
the ratio of gap frequency over critical temperature $\omega_{g}/T_{c}$
is far less than the value $8$ which is found in the standard version holographic superconductors\cite{3H,Hvarious},
while for small $\gamma$, the value of the ratio is larger than $8$.
Therefore, it is worth exploring how the picture of the St\"{u}ckelberg holographic superconductor is modified
when considering Weyl corrections.

Our paper is organized as follows. In section II,
we construct St\"{u}ckelberg holographic superconductor with Weyl corrections and present numerical results
from the St\"{u}ckelberg holographic superconductor.
We specially discuss the order of phase transitions
and the critical exponents of second order phase transitions.
Then we investigate the conductivity numerically in section III.
Conclusions and discussions are followed in section IV.

\section {The St\"{u}ckelberg holographic superconductor with Weyl corrections}

\subsection{The St\"{u}ckelberg holographic superconductor with Weyl corrections}

The general St\"{u}ckelberg holographic superconductor contains
the gravitational field, the $U(1)$ gauge field and the scalar field, the latter two of which are
coupled via a general St\"{u}ckelberg mechanics \cite{stuckelberg2}.
The action in this system is
\begin{eqnarray}
\label{stckelbergaction}
S=\int d^{d+1}x \sqrt{-g} \left( R + \frac{(d-1)d}{L^{2}} -\mathcal{L} \right),
\end{eqnarray}
with
\begin{eqnarray}
\label{stckelbergL}
\mathcal{L}=\frac{F_{\mu\nu}F^{\mu\nu}}{4}
+\frac{\partial_{\mu}\tilde{\Psi}\partial^{\mu}\tilde{\Psi}}{2}
+\frac{m^{2}}{2}\tilde{\Psi}^{2}
+\frac{1}{2}\mathcal{F}(\tilde{\Psi})(\partial_{\mu}p-A_{\mu})(\partial^{\mu}p-A^{\mu}),
\end{eqnarray}
where $\mathcal{F}$ is a general function of $\tilde{\Psi}$
\begin{eqnarray}
\label{functionofF}
\mathcal{F}(\tilde{\Psi})=\tilde{\Psi}^{2}+c_{\alpha}\tilde{\Psi}^{\alpha}+c_{4}\tilde{\Psi}^{4},
\end{eqnarray}
in which $c_{4}$, $c_{\alpha}$, $\alpha$ are real numbers and $3 \leq \alpha \leq 4$.
When $c_{\alpha} = c_{4} = 0$, it reduces to the model of \cite{3H}.
In addition, this theory has the local gauge symmetry $A_{\mu}\rightarrow A_{\mu}+\partial_{\mu}\Omega(x^{\mu})$ and $p\rightarrow p+\Omega(x^{\mu})$,
where we can use the gauge freedom to fix $p=0$.
Furthermore, we take the ansatz $\tilde{\Psi}=\Psi$.
Therefore, the St\"{u}ckelberg Lagrangian (\ref{stckelbergL}) can be re-expressed as
\begin{eqnarray}
\label{stckelbergL1}
\mathcal{L}=\frac{F_{\mu\nu}F^{\mu\nu}}{4}
+\frac{\partial_{\mu}\Psi\partial^{\mu}\Psi}{2}
+\frac{m^{2}}{2}\Psi^{2}
+\frac{1}{2}\mathcal{F}(\Psi)A_{\mu}A^{\mu}.
\end{eqnarray}

The function $\mathcal{F}$ becomes
\begin{eqnarray}
\label{functionofF1}
\mathcal{F}(\Psi)=\Psi^{2}+c_{\alpha}\Psi^{\alpha}+c_{4}\Psi^{4}.
\end{eqnarray}

In this paper, we will construct a $3+1$ dimension St\"{u}ckelberg holographic superconductor
with Weyl corrections as Ref.\cite{JPWu}.
In this case, the matter Lagrangian density $\mathcal{L}$
is replaced by
\begin{eqnarray}
\label{matterLWeyl}
\mathcal{L}_{W}=\frac{1}{4}\left(F^{\mu\nu} F_{\mu\nu}-4\gamma C^{\mu\nu\rho\sigma}F_{\mu\nu}F_{\rho\sigma}\right)
+\frac{\partial_{\mu}\Psi\partial^{\mu}\Psi}{2}
+\frac{m^{2}}{2}\Psi^{2}
+\frac{1}{2}\mathcal{F}(\Psi)A_{\mu}A^{\mu},
\end{eqnarray}
where $\gamma$ is a (dimensionful) constant and $C_{\mu\nu\rho\sigma}$ is the Weyl tensor.
The range of the parameter $\gamma$ \cite{WeylM} is
\begin{eqnarray}
\label{gamma}
-\frac{L^{2}}{16}<\gamma<\frac{L^{2}}{24},
\end{eqnarray}
where the upper bound results from the existence of an additional singular
point at $\gamma=\frac{L^{2}}{24}$ and the lower bound is due to the causality constraints.

We exclusively work in the probe approximation in this letter,
where the gravity sector is effectively decoupled from the matter sector.
So we take the background to be a planar AdS-Schwarzschild black hole:
\begin{eqnarray}
\label{metric1}
ds^{2}=\frac{r^{2}}{L^{2}}(-f(r)dt^{2}+dx_{i}dx^{i})+\frac{L^{2}}{r^{2}f(r)}dr^{2},
\end{eqnarray}
where
\begin{eqnarray}
\label{f1}
f(r)=1-\left(\frac{r_{H}}{r}\right)^{4},
\end{eqnarray}
with $r_{H}>0$ for the horizon and $r\rightarrow \infty$ for boundary of the bulk,
the Hawking temperature of this black hole is
\begin{eqnarray}
\label{HawkingT}
T=\frac{r_{H}}{\pi L^{2}},
\end{eqnarray}
which corresponds to the temperature of the dual field theory.
For convenience, in the following analysis, we always set $L=1$.
Let's begin with the following Ansatz:
\begin{eqnarray}
\label{ansatz}
\Psi=\Psi(r),~~~ A_{t}=\Phi(r),
\end{eqnarray}
then apply it to the Euler-Lagrange equation,
we derive the equations of motion from the Weyl corrections matter Lagrangian density (\ref{matterLWeyl})
\begin{eqnarray}
\label{ScalarEOM1}
\Psi''+\left(\frac{f'}{f}+\frac{5}{r}\right)\Psi'+\frac{\Phi^{2}}{ 2 r^{4} f^{2}} \dot{\mathcal{F}} -\frac{m^{2}\Psi}{r^{2}f}=0,
\end{eqnarray}
\begin{eqnarray}
\label{GMaxwellEOM1}
\left(1-\frac{24\gamma r_{H}^{4}}{r^{4}}\right)\Phi''
+\left(\frac{3}{r}+\frac{24\gamma r_{H}^{4}}{r^{5}}\right)\Phi'
-\frac{\mathcal{F}}{r^{2} f} \Phi
=0,
\end{eqnarray}
where the prime represents the derivative with respect to $r$
and the dot in $\mathcal{F}$ represents the derivative with respect to $\Psi$.
In addition, we have used the nonzero components of Weyl tensor $C_{\mu\nu\rho\sigma}$ in $AdS_{5}$,
\begin{eqnarray}
\label{WeylT}
C_{0i0j}=f(r)r_{H}^{4} \delta_{ij},~~
C_{0r0r}=-\frac{3 r_{H}^{4}}{r^{4}},~~
C_{irjr}=-\frac{r_{H}^{4}}{r^{4} f(r)} \delta_{ij},~~
C_{ijkl}=r_{H}^{4} \delta_{ik} \delta_{jl}.
\end{eqnarray}

For concreteness, we fix the mass of the scalar field to $m^{2}=-3$,
which is above the Breitenlohner-Freedman bound \cite{BF}
and does not induce an instability.
With this mass, the asymptotic behavior of the scalar fields $\Psi$
and the scalar potential $\Phi$
close to the AdS boundary ($r\rightarrow \infty$) take the form
\begin{eqnarray}
\label{BconditionI1}
\Psi=\frac{\Psi_{1}}{r}+\frac{\Psi_{3}}{r^{3}},
\end{eqnarray}
\begin{eqnarray}
\label{BconditionI2}
\Phi=\mu-\frac{\rho}{r^{2}},
\end{eqnarray}
where $\mu$ and $\rho$ are the chemical potential and charge density
in the dual field theory, respectively.
For the value of $m^{2}$ we have mentioned before, only the $\Psi_{3}$ term is normalized.
Here we set $\Psi_{1}=0$, so the expectation value of the scalar operator
$\mathcal{O}_{3}$ in the dual field theory
is related to $\Psi_{3}$ by $\langle\mathcal{O}_{3}\rangle=\Psi_{3}$.
Furthermore, at the horizon $r=r_{H}$, the regularity implies
\begin{eqnarray}
\label{BconditionH1}
\Psi'(r_{H})=-\frac{3}{4 r_{H}}\Psi(r_{H}),
\end{eqnarray}
\begin{eqnarray}
\label{BconditionH2}
\Phi(r_{H})=0.
\end{eqnarray}

To investigate the behavior of phase transitions of this system, we solve equations of motion (\ref{ScalarEOM1}) and (\ref{GMaxwellEOM1}) combined with the boundary conditions (\ref{BconditionI1}), (\ref{BconditionI2}), (\ref{BconditionH1}) and (\ref{BconditionH2}) numerically by using a shooting method.

\subsection{Phase transitions}

One main characteristic of the St\"{u}ckelberg holographic superconductor is that
it enables us to describe various types of phase transitions, from first to second order. As for the second one, it is also able to illustrate those with the mean and non-mean field behaviors.
When involving the Weyl corrections, how the phase transition depends on the coefficients $c_{\alpha}$, $\alpha$, $c_{4}$ and Weyl coupling parameter $\gamma$ is deeply investigated.
In order to see the role each parameter plays, we observe the following three cases respectively.

\subsubsection{The case $\mathcal{F}(\Psi)=\Psi^{2}+c_{4}\Psi^{4}$}

\begin{figure}
\center{
\includegraphics[scale=0.95]{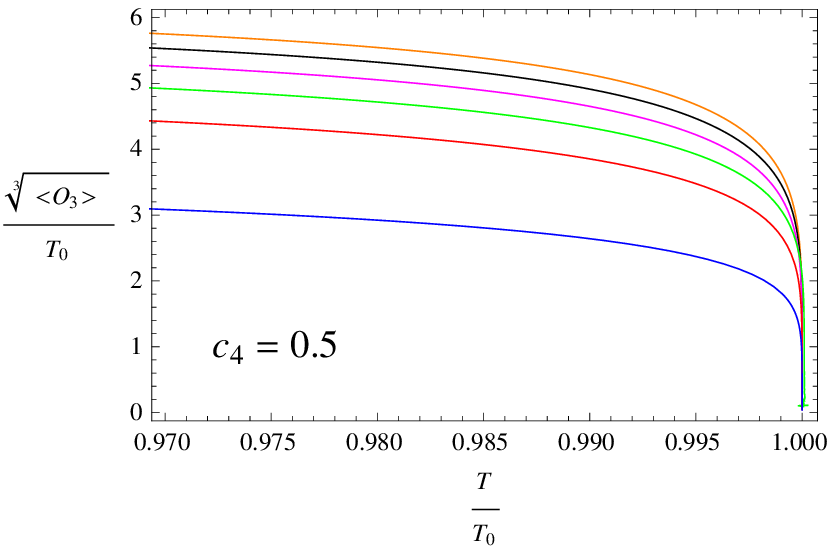}\hspace{0.05cm}
\includegraphics[scale=0.95]{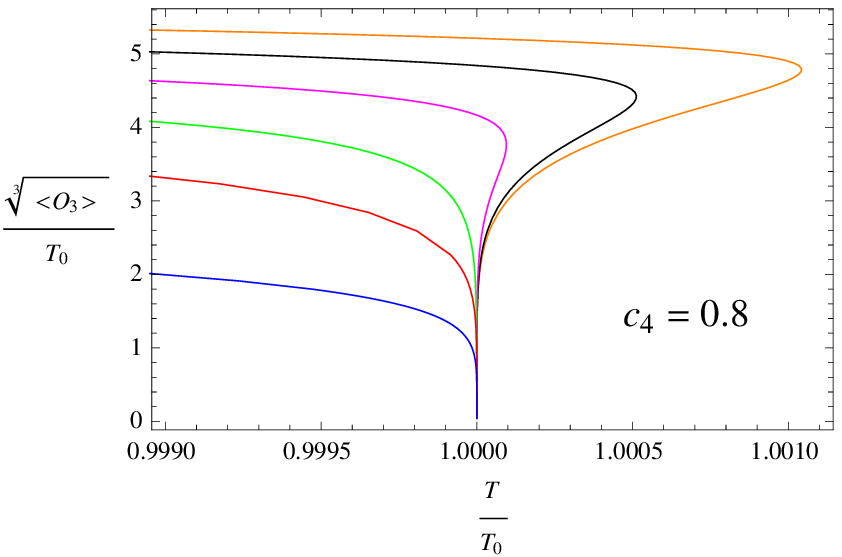}\\ \hspace{0.05cm}}
\center{
\includegraphics[scale=0.95]{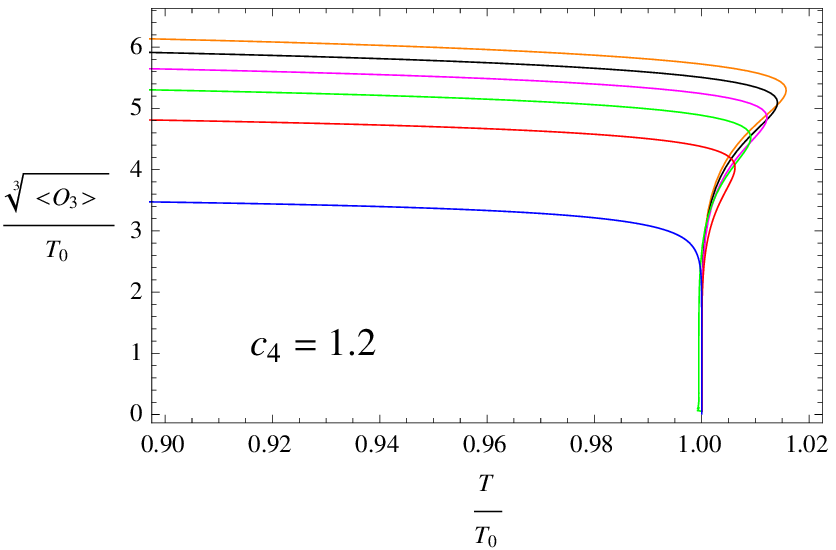}\hspace{0.05cm}
\includegraphics[scale=0.95]{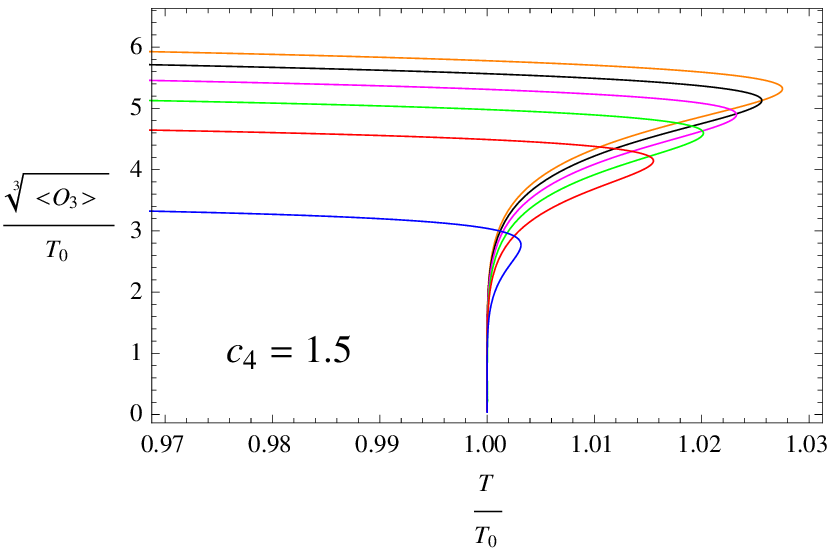}\\ \hspace{0.05cm}
\caption{\label{condensationc4} The condensate around the critical region for chosen values of $c_{4}$
and different values of $\gamma$. $T_{0}$ is the temperature when condensate
and second order phase transition appear, $T_{0}$ equals to the critical temperature $T_{c}$.
The orange, black, magenta, green, red and blue curves correspond to
$\gamma=-0.06, -0.04, -0.02, 0, 0.02, 0.04$, respectively.}}
\end{figure}

Firstly, we focus on the special case $\mathcal{F}(\Psi)=\Psi^{2}+c_{4}\Psi^{4}$
to see how $c_{4}$ and $\gamma$ control the styles of the phase transition.
Solving the equations of motion (\ref{ScalarEOM1}) and (\ref{GMaxwellEOM1}) numerically,
we have the results itemized as follow: over the legal range of $\gamma$,
the phase transition is second order when $0\leq c_{4}\lesssim 0.7$
and is first order for $c_{4} \gtrsim 1.3$ (Fig.\ref{condensationc4}).
However, if $0.8 \lesssim c_{4}\lesssim 1.2$, the phase transition varies from second order to first order, and first order ones are prone to appear in smaller $\gamma$, while the second order ones usually emerge in larger $\gamma$ (Fig.\ref{condensationc4}).
Furthermore, by fixing the data, we find that for the second order phase transition, while for $T\sim T_{c}$,
there is a relationship between $\mathcal{O}_{3}$ and $T_{c}$,
that is, $\mathcal{O}_{3}\sim (T_{c}-T)^{\beta}$,
 $\beta=1/2$, which has nothing to do with $\gamma$.

\subsubsection{The case $\mathcal{F}(\Psi)=\Psi^{2}+c_{\alpha}\Psi^{\alpha}$}

\begin{figure}
\center{
\includegraphics[scale=0.95]{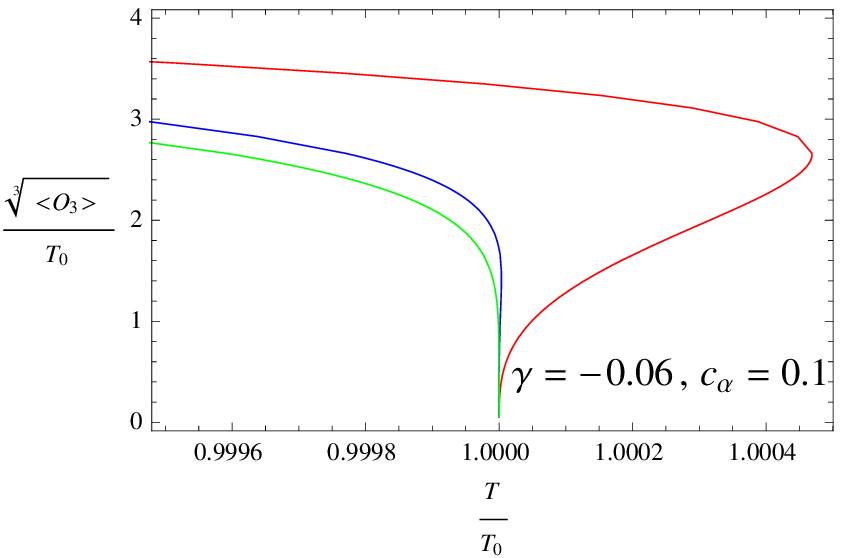}\hspace{0.05cm}
\includegraphics[scale=0.95]{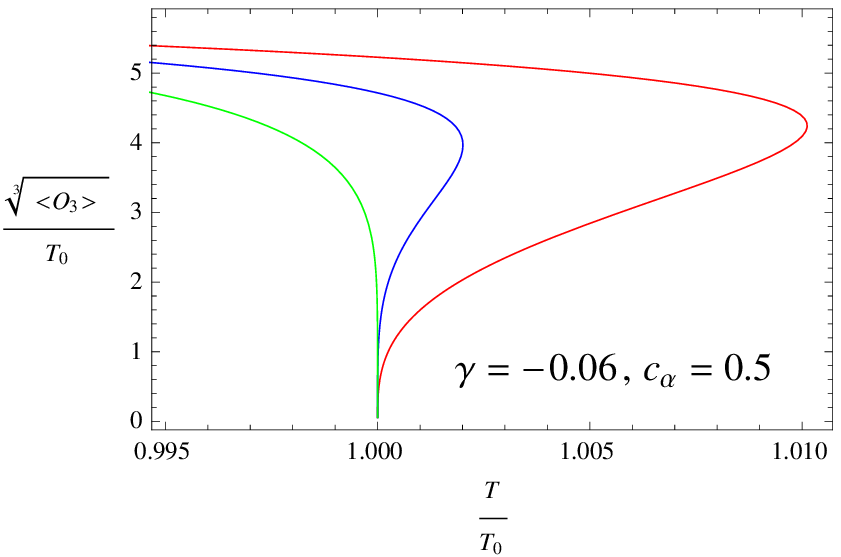}\\ \hspace{0.05cm}}
\center{
\includegraphics[scale=0.95]{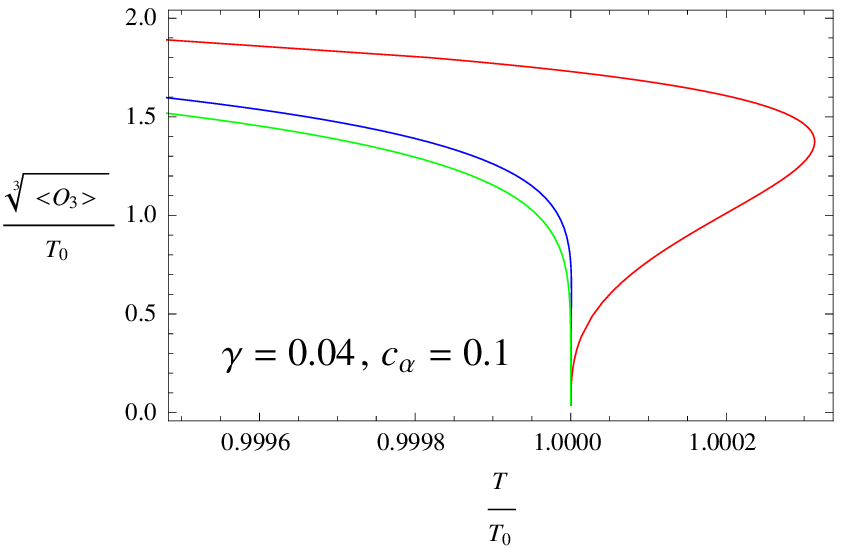}\hspace{0.05cm}
\includegraphics[scale=0.95]{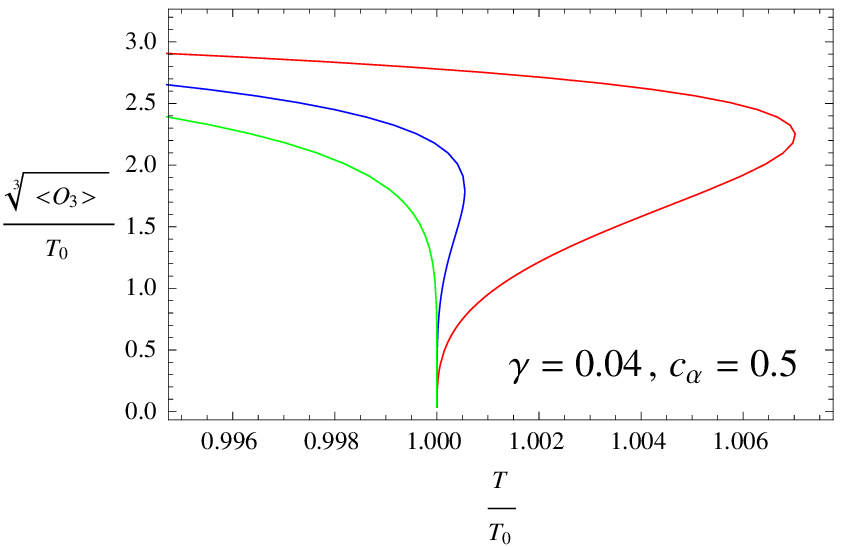}\\ \hspace{0.05cm}
\caption{\label{condensationcalpha} The condensate around the critical region
for chosen values of $\gamma$, $c_{\alpha}$ and different values of $\alpha$.
The red, blue and green curves correspond to $\alpha =3, 3.5, 4$, respectively.}}
\end{figure}

Now, in order to investigate the influence of $c_{\alpha}$, $\alpha$ and $\gamma$ on the phase transition, we set $c_{4}=0$. In Fig.\ref{condensationcalpha}, we present the condensate around the critical region for chosen values of $\gamma$, $c_{\alpha}$ and different values of $\alpha$.
Fig.\ref{condensationcalpha} informs us that for chosen values of $\gamma$ and $c_{\alpha}$,
first order phase transitions often gather in the region of smaller $\alpha$, whereas second order ones usually lie in the region of larger $\alpha$.
Moreover, for fixed $\gamma$ and $\alpha$, the styles of phase transition exhibit some kinds of similar distribution, but what is different is that first order ones appear with larger $c_{\alpha}$, while second order ones happen when $c_{\alpha}$ is small.
To sum up, more often than not, the first order phase transition locates in the region of larger $c_{\alpha}$ or smaller $\alpha$,
which is consistent with the result in Gauss-Bonnet gravity \cite{stuckelberg4}.
Then, we will specially concentrate on the effect of the Weyl parameter $\gamma$ on the phase transition.
From Fig.\ref{condensationcalpha}, we find that for chosen values of $\alpha$ and $c_{\alpha}$, first order phase transitions are distributed in small $\gamma$, while second order ones occupy the larger $\gamma$ region.
In light of discussions above, one can conclude that the first order phase transition of the system is liable to emerge when Weyl parameter
$\gamma$ is small.

\subsubsection{The case $\mathcal{F}(\Psi)=-\Psi^{\alpha}+\frac{1}{2}\Psi^{4}$}

\begin{figure}
\center{
\includegraphics[scale=0.95]{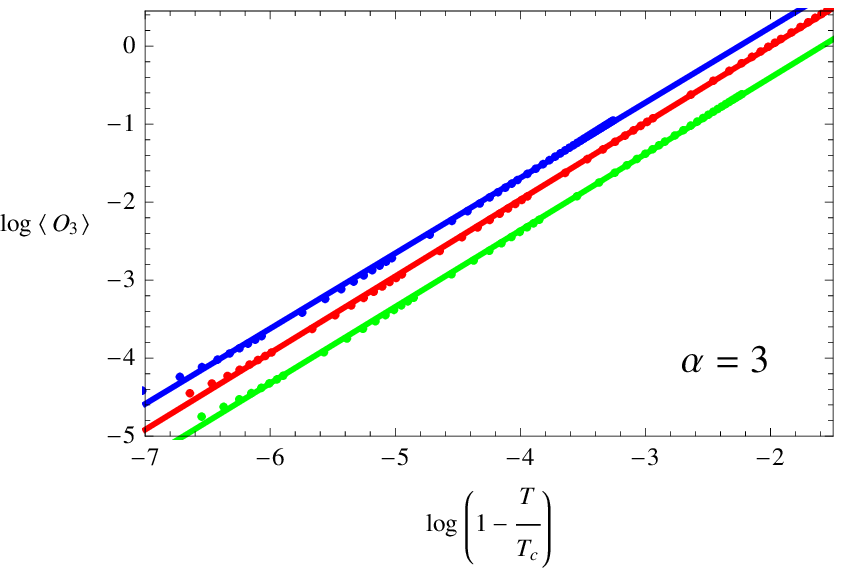}\hspace{0.05cm}
\includegraphics[scale=0.95]{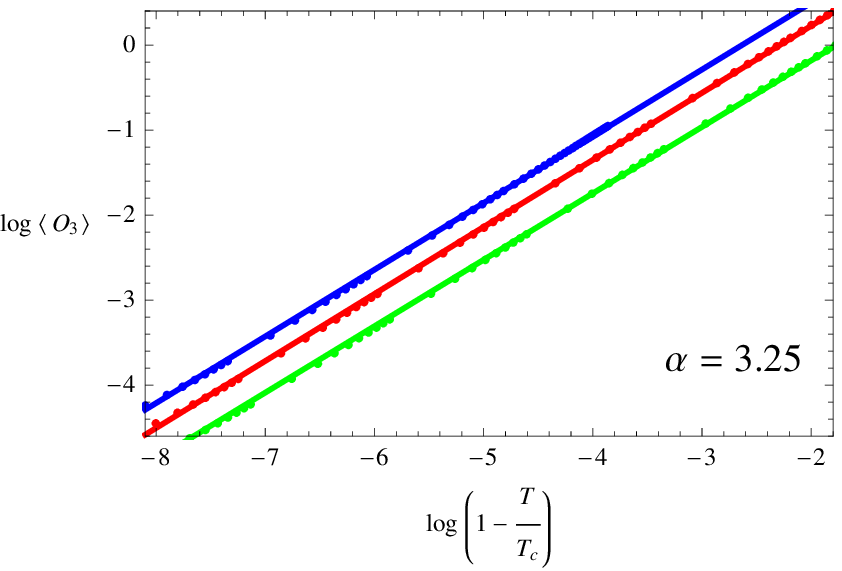}\\ \hspace{0.05cm}}
\center{
\includegraphics[scale=0.95]{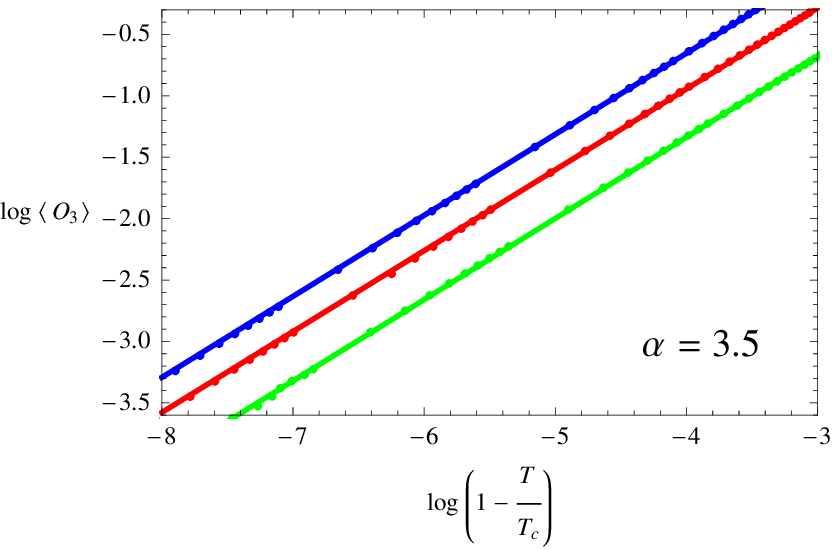}\hspace{0.05cm}
\includegraphics[scale=0.95]{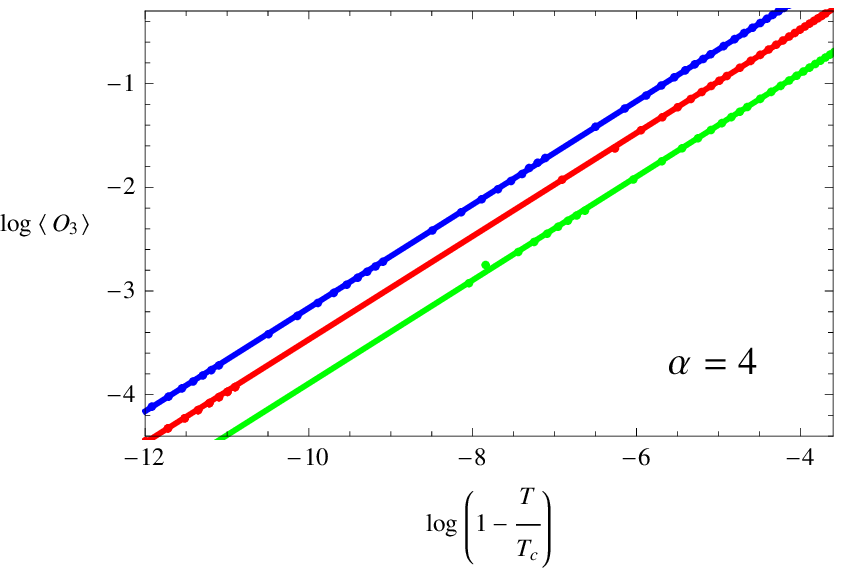}\\ \hspace{0.05cm}
\caption{\label{condensafit} The critical exponent $\beta$ (the slope of the curves)
is independent on the the Weyl parameter $\gamma$ and only dependent on $\alpha$.
The vertical axis is the log of the Condensate $\langle \mathcal{O}_{3}\rangle$ for the case $\mathcal{F}(\Psi)=-\Psi^{\alpha}+\frac{1}{2}\Psi^{4}$
near the critical temperature $T_{c}$ and the horizontal axis is the log of $1-T/T_{c}$.
The blue, red and green curves correspond to $\gamma =-0.06, 0.02, 0.04$, respectively.}}
\end{figure}

In the above two cases, we have shown that different model parameters can lead to either first
or second order phase transitions.
It is natural to ask whether the change of model parameters would exert some influence on the critical exponent, rendering it different from the prediction of the mean filed. In reality, the answer is yes. Ref.\cite{stuckelberg2} has shown that for $c_{\alpha}< 0$, the critical exponent is not the same as the result the mean field holds when model parameters are varied.
When involving the Weyl corrections, we also focus on
the special case $c_{\alpha}=-1$ and $c_{4}=1/2$ as in Ref.\cite{stuckelberg2}.
From Fig.\ref{condensafit}, we find that the critical exponent $\beta$
is independent of the Weyl parameter $\gamma$ but only dependent on $\alpha$.
Adopting  the non-linear fitting method, we can get the critical exponent $\beta$ for different $\alpha$.
Therefore, the relation
$\beta=(\alpha-2)^{-1}$ found in Ref.\cite{stuckelberg2} is still preserved.
These results show that the critical exponent $\beta$ depends only on the model parameter $\alpha$
but is independent of the Weyl parameter $\gamma$. Actually, according to \cite{stuckelberg2,stuckelberg4}, other parameters, such as the scalar mass, the dimensions of space-times
and the background geometry, have no influence on the value of the critical exponent $\beta$. Thus, we can say that in this certain case $\mathcal{F}(\Psi)=-\Psi^{\alpha}+\frac{1}{2}\Psi^{4}$, the critical exponent is decided by parameter $\alpha$.

\begin{widetext}
\begin{table}[ht]
\begin{center}

\begin{tabular}{|c|c|c|c|c|c|c|}
         \hline
$~~\alpha~~$ &~~$3$~~&~~$3.25$~~&~~$3.5$~~&~~$4$~~
         \\
        \hline
~~$\beta$~~ & ~~$1$~~ & ~~$0.79$~~ & ~~$0.66$~~& ~~$0.5$~~
          \\
        \hline
\end{tabular}
\caption{\label{beta} The critical exponent $\beta$ for the case $\mathcal{F}(\Psi)=-\Psi^{\alpha}+\frac{1}{2}\Psi^{4}$ with different $\alpha$.}

\end{center}
\end{table}
\end{widetext}

\begin{figure}
\center{
\includegraphics[scale=0.95]{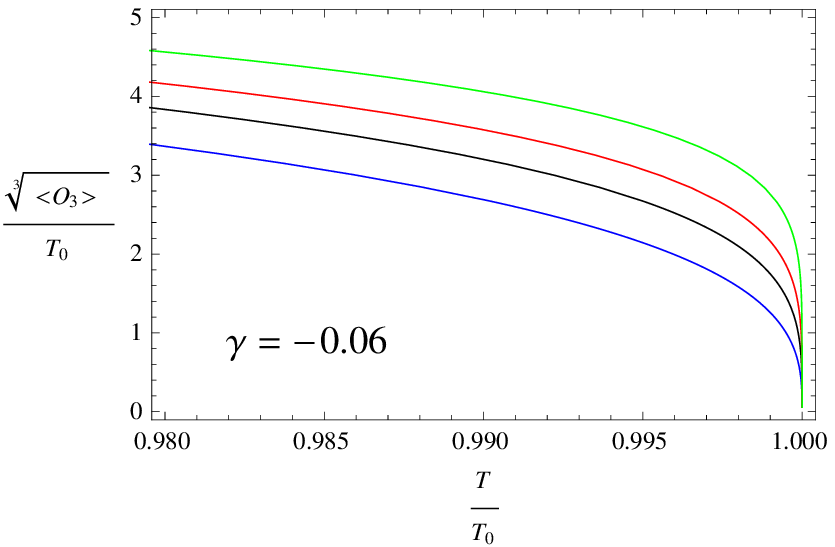}\hspace{0.05cm}
\includegraphics[scale=0.95]{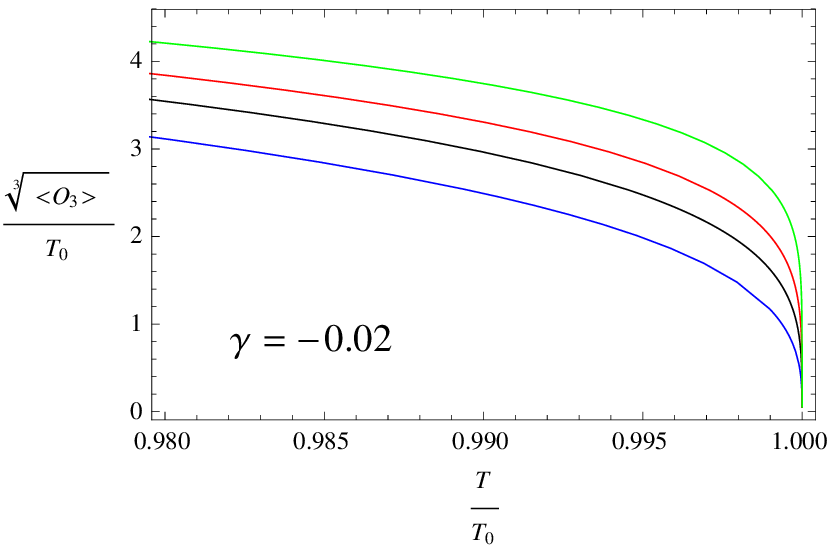}\\ \hspace{0.05cm}}
\center{
\includegraphics[scale=0.95]{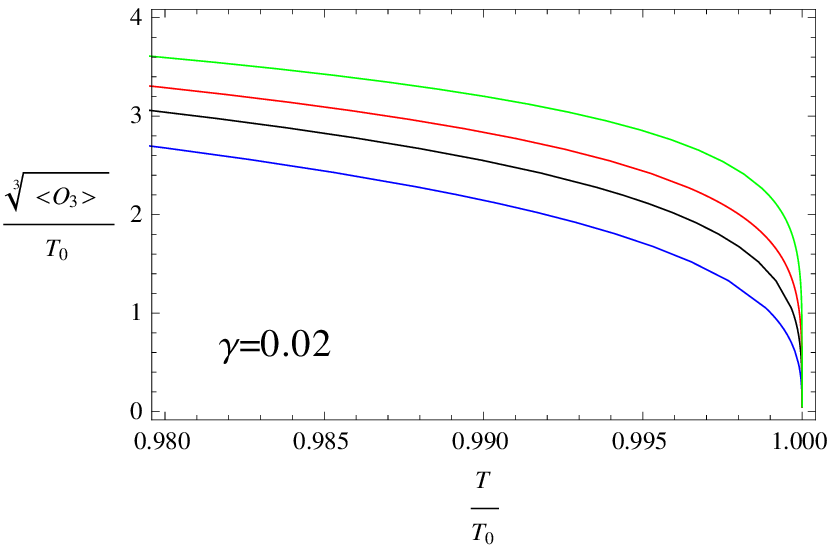}\hspace{0.05cm}
\includegraphics[scale=0.95]{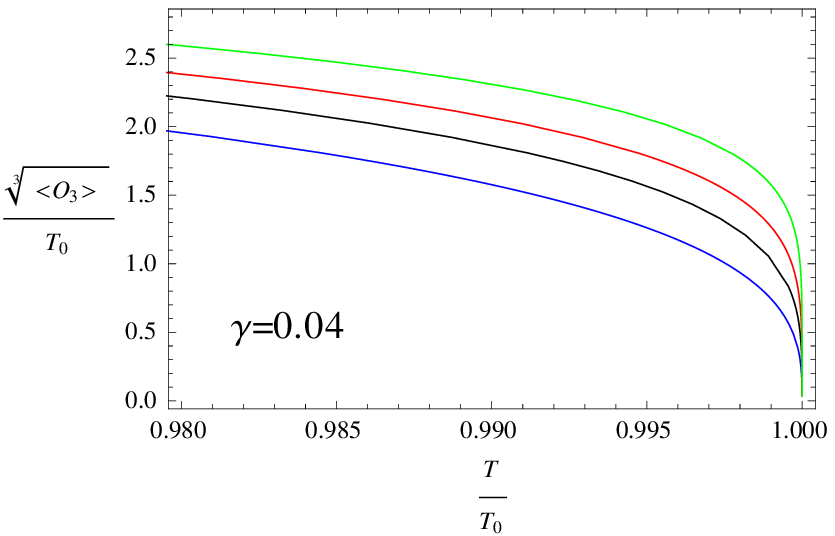}\\ \hspace{0.05cm}
\caption{\label{condensationcother} The condensate around the critical region for chosen values of $\gamma$, $c_{\alpha}$
and different values of $\alpha$.
The red, blue and green curves correspond to
$\alpha =3, 3.5, 4$, respectively.}}
\end{figure}

Next, we will focus on the dependence of the condensate on $\alpha$ and $\gamma$.
First, we present the condensate as a function of temperature in Fig.\ref{condensationcother}.
From these figures above, it is easy to see that the given different values of $\alpha$ and $\gamma$, the critical temperature $T_{c}$ are unrelated to $\alpha$ and only rely on $\gamma$.
In the meantime, the critical temperature $T_{c}$ for different Weyl parameter $\gamma$ is summarized in Table \ref{Tctable}.
Observing this table, we claim that the critical temperatures $T_{c}$ is higher and higher
as the Weyl parameter $\gamma$ changes from $-0.06$ to $0.04$, which is in agreement with Ref.\cite{JPWu}.
In another word, when Weyl parameter $\gamma$ runs toward the negative direction, it is much harder for the scalar hair to form.
However, if $\gamma$ turns around, the formation of the scalar hair would be facilitated.
In fact, for the second order phase transition in all cases above, the critical temperature $T_{c}$ only
depends on the Weyl parameter and the form of $\mathcal{F}(\Psi)$ is not the determinant of $T_{c}$.

\begin{widetext}
\begin{table}[ht]
\begin{center}

\begin{tabular}{|c|c|c|c|c|c|c|}
         \hline
$~~\gamma~~$ &~~$-0.06$~~&~~$-0.02$~~&~~$0.02$~~&~~$0.04$~~
         \\
        \hline
~~$T_{c}$~~ & ~~$0.170\rho^{1/3}$~~ & ~~$0.185\rho^{1/3}$~~ & ~~$0.219\rho^{1/3}$~~& ~~$0.304\rho^{1/3}$~~
          \\
        \hline
\end{tabular}
\caption{\label{Tctable} The critical temperature $T_{c}$ for the case $\mathcal{F}(\Psi)=-\Psi^{\alpha}+\frac{1}{2}\Psi^{4}$ with different $\gamma$.}

\end{center}
\end{table}
\end{widetext}

\section {Electrical conductivity}

\begin{figure}
\center{
\includegraphics[scale=0.62]{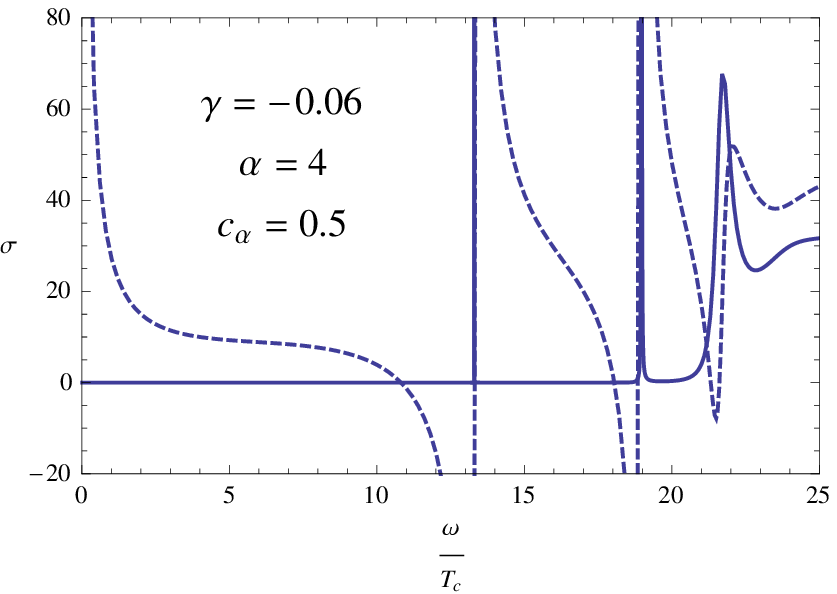}\hspace{0.05cm}
\includegraphics[scale=0.62]{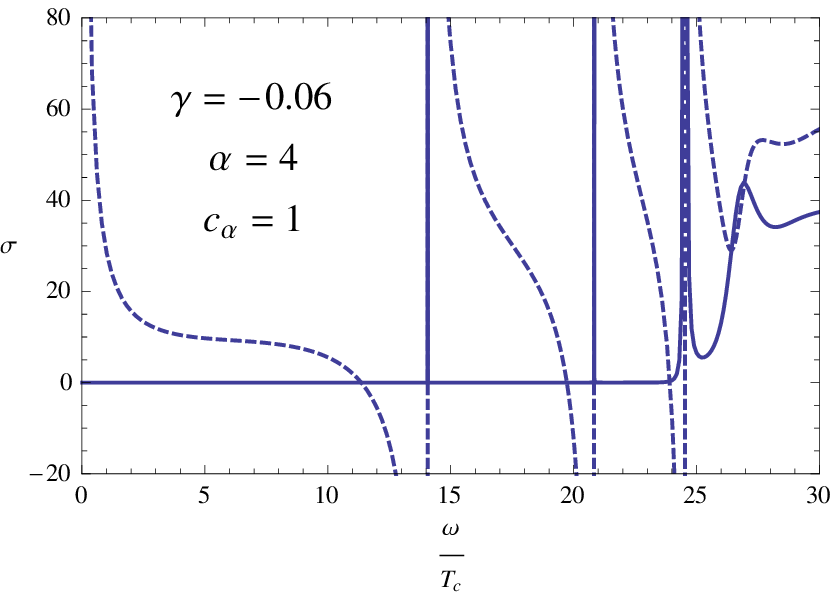}\hspace{0.05cm}
\includegraphics[scale=0.62]{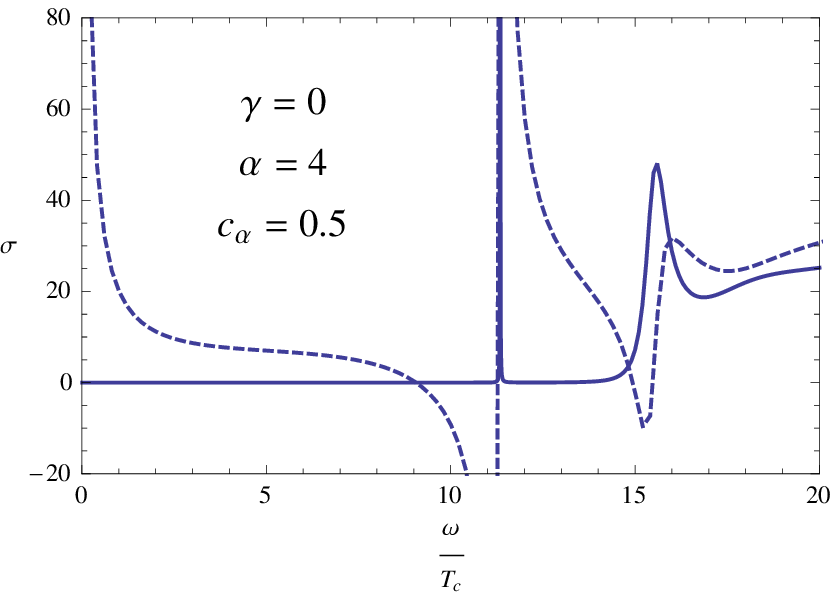}\\ \hspace{0.05cm}}
\center{
\includegraphics[scale=0.62]{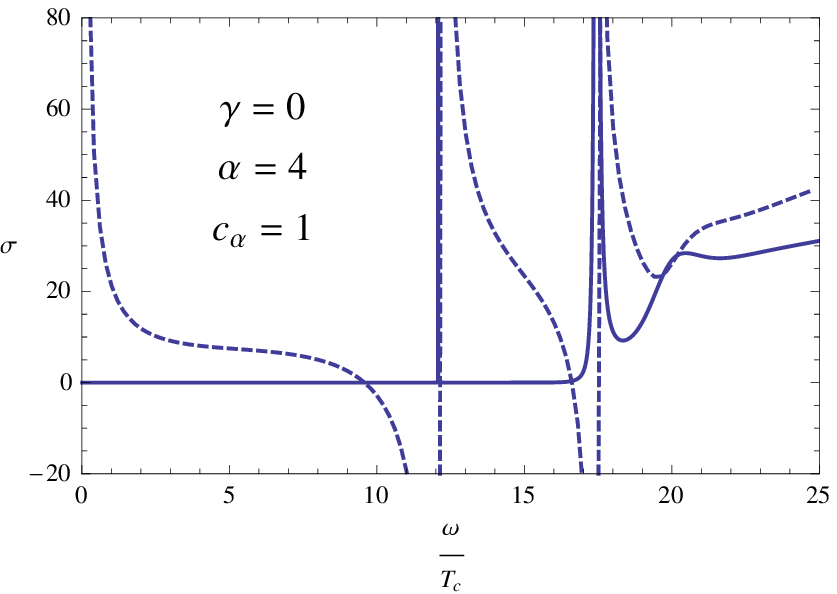}\hspace{0.05cm}
\includegraphics[scale=0.62]{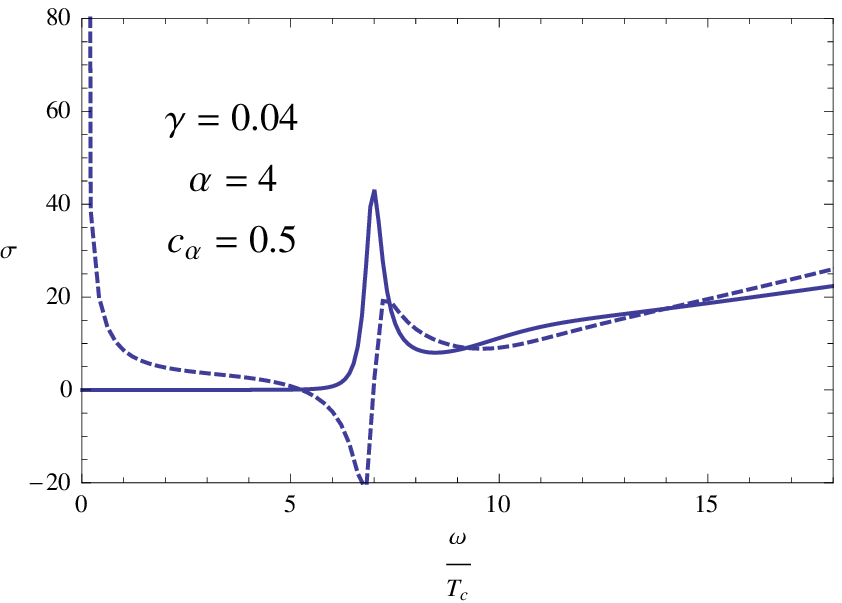}\hspace{0.05cm}
\includegraphics[scale=0.62]{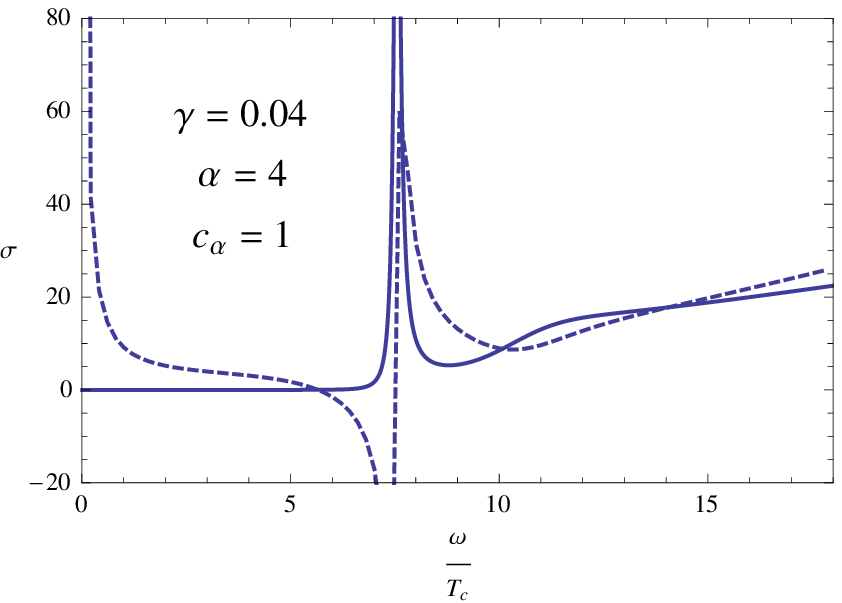}\\ \hspace{0.05cm}}
\center{
\includegraphics[scale=0.62]{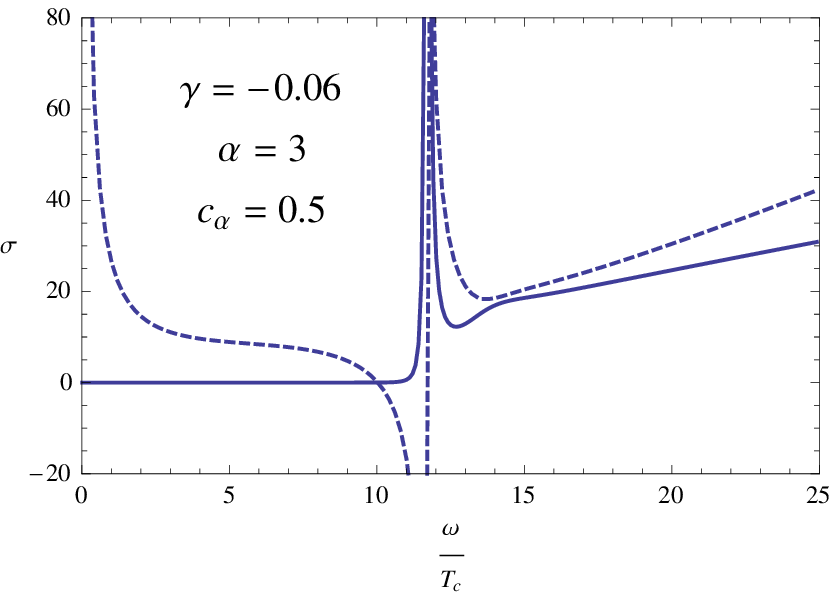}\hspace{0.05cm}
\includegraphics[scale=0.62]{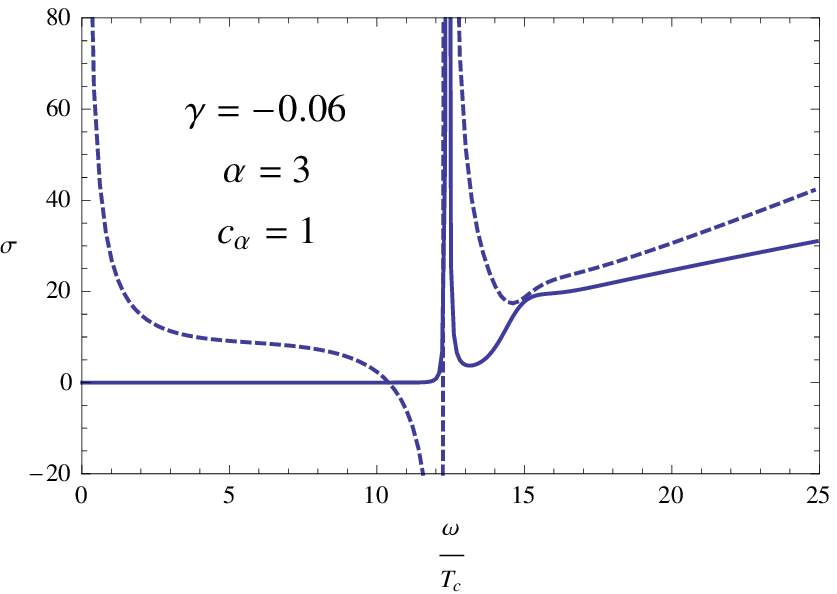}\hspace{0.05cm}
\includegraphics[scale=0.62]{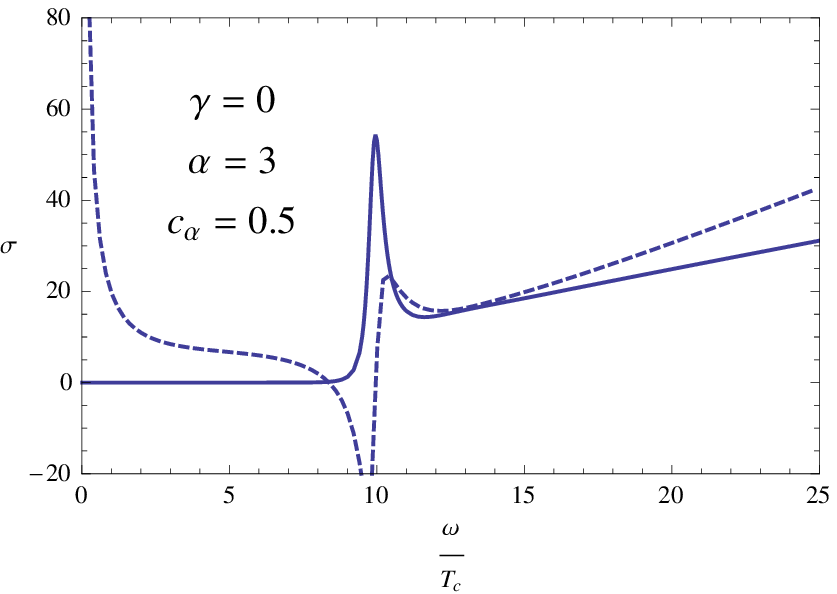}\\ \hspace{0.05cm}}
\center{
\includegraphics[scale=0.62]{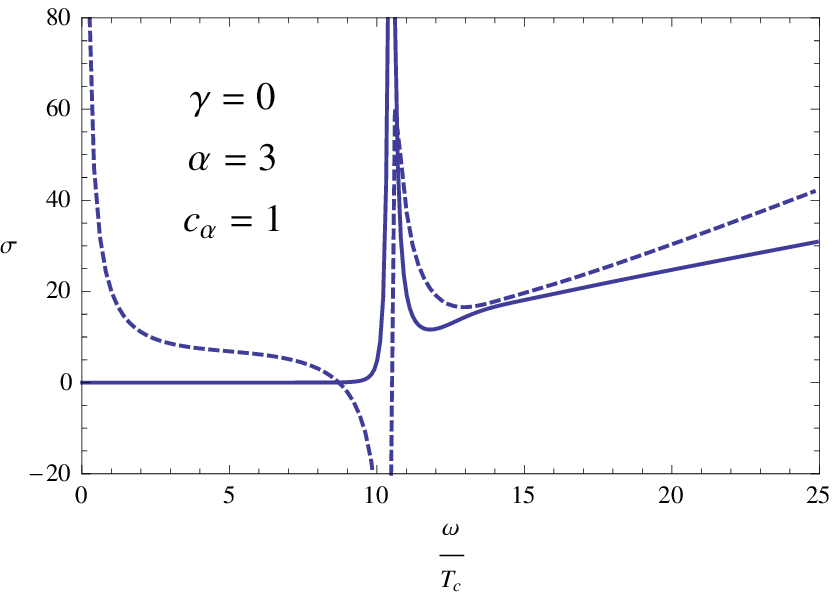}\hspace{0.05cm}
\includegraphics[scale=0.62]{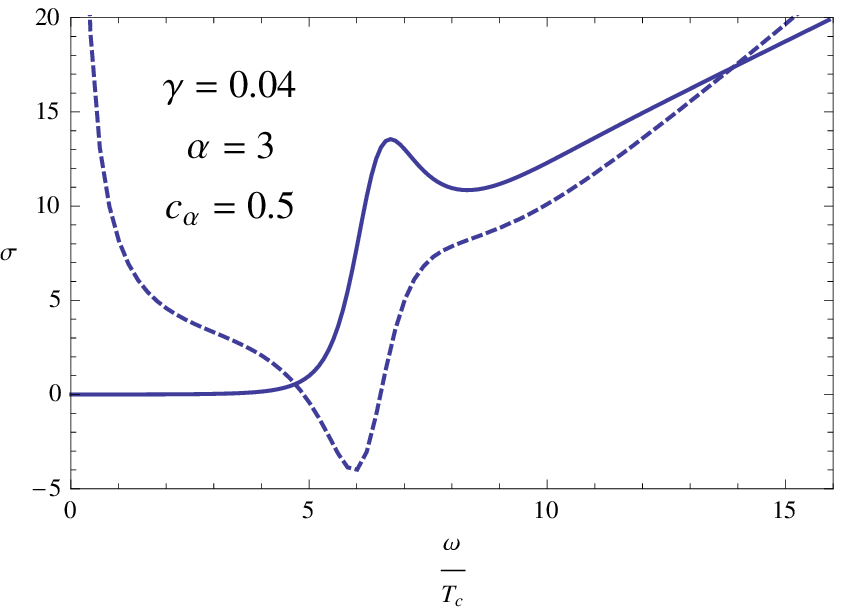}\hspace{0.05cm}
\includegraphics[scale=0.62]{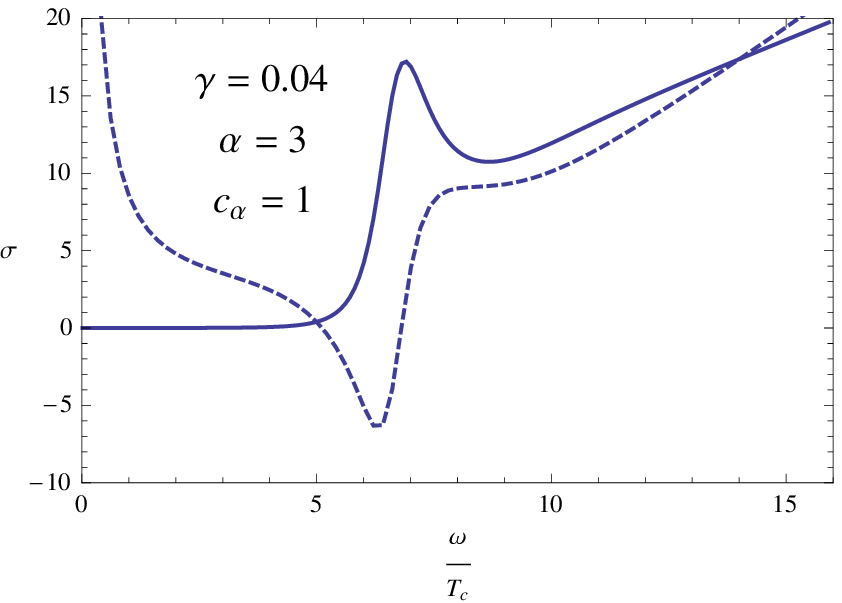}\\ \hspace{0.05cm}
\caption{\label{conductivity1} Conductivity for superconductors with different values
of $\alpha$, $c_{\alpha}$ and $\gamma$ at the same temperature $T/T_{c}\approx 0.4$.
The solid line represents the real part, and dotted line the imaginary part.}}
\end{figure}

\begin{figure}
\center{
\includegraphics[scale=0.62]{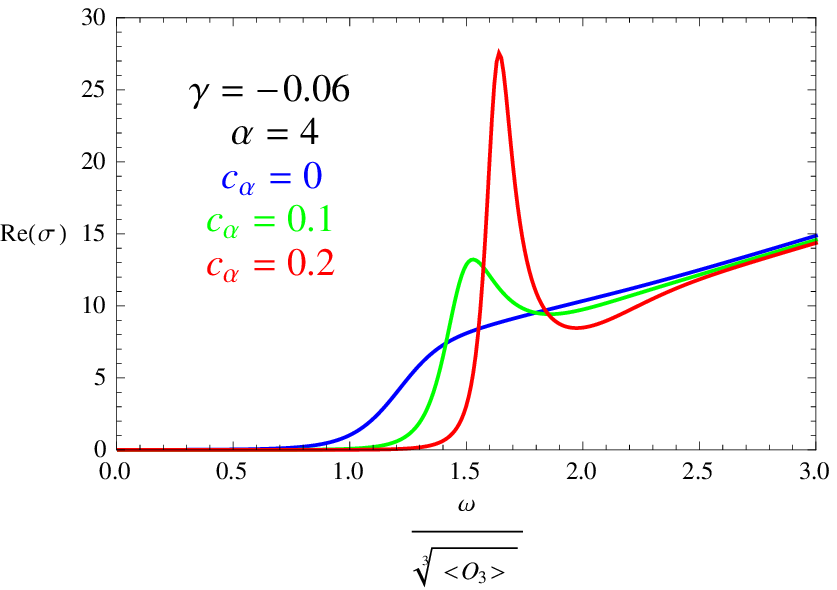}\hspace{0.05cm}
\includegraphics[scale=0.62]{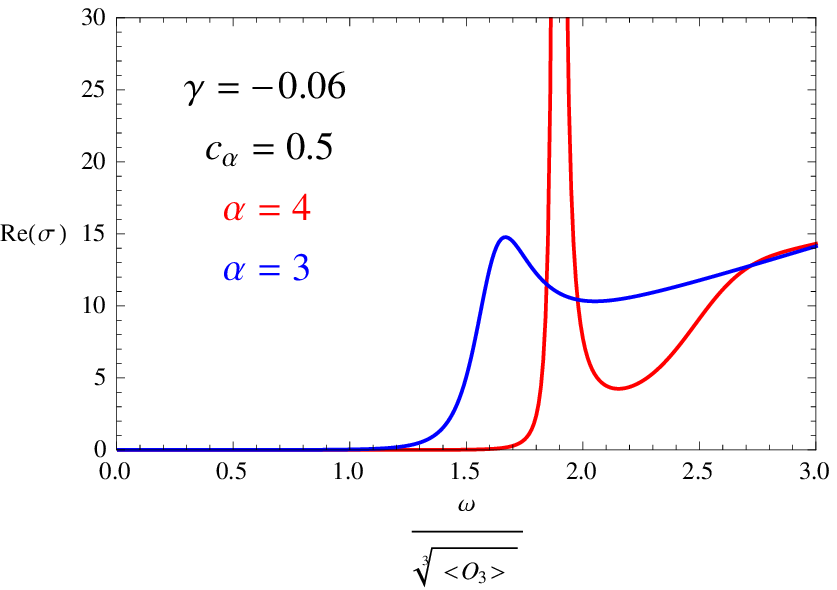}\hspace{0.05cm}
\includegraphics[scale=0.62]{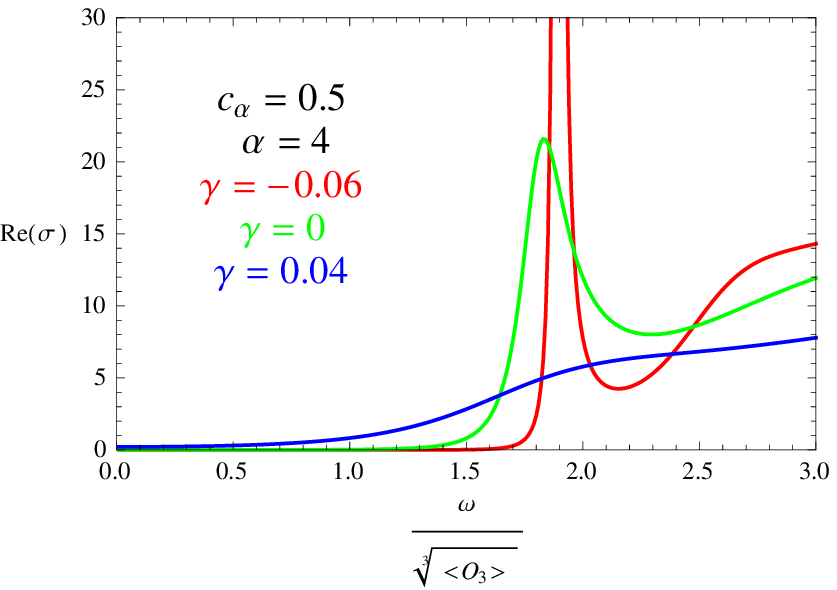}\\ \hspace{0.05cm}
\caption{\label{conductivity2} $Re(\sigma)$ as a function of $\omega$ in
units of the condensate with different values of $\alpha$, $c_{\alpha}$
and $\gamma$ at the same temperature $T/T_{c}\approx 0.7$.
}}
\end{figure}

In this section, we study the electrical conductivity $\sigma$ of the dual field theory.
We switch on a vector perturbation of radial symmetry and time dependence in the $x$-direction,
$\delta A_{x}(t,r)=A_{x}(r)e^{-i\omega t}dx$ \footnote{for simplicity, we will work at zero momentum ($k=0$).},
and study the linear response of the system.
With the help of the Euler-Lagrange equation,
the equation of motion for the perturbation of vector potential $\delta A_{x}(t,r)$ is obtained in the following
\begin{eqnarray}
\label{GMaxwellxEOM}
(1&+&\frac{8\gamma r_{H}^{4}}{r^{4}})A''_{x}
+\left[\left(\frac{f'}{f}+\frac{3}{r}\right)+ \frac{8\gamma r_{H}^{4}}{r^{4}}\left(\frac{f'}{f}-\frac{1}{r}\right)\right]A'_{x}
\nonumber\\
&& \quad
+\left(1+\frac{8\gamma r_{H}^{4}}{r^{4}}\right)\frac{\omega^{2}}{r^{4}f^{2}}A_{x}
-\frac{\mathcal{F}(\Psi)}{r^{2} f}A_{x}
=0.
\end{eqnarray}

At the horizon, the ingoing boundary conditions due to the requirement of causal behavior \cite{Son} must be imposed.
In our case, this is
\begin{eqnarray}
\label{ingoingC}
A_{x}(r)\sim (r^{2}f(r))^{-i\frac{\omega}{4r_{H}}},
\end{eqnarray}

Around the AdS boundary $r\rightarrow \infty$, $A_{x}(r)$ is expanded as
\begin{eqnarray}
\label{Axinfty}
A_{x}(r)=A^{(0)}+\frac{A^{(2)}}{r^{2}}+\frac{A^{(0)}\omega^{2}}{2}\frac{\log \Lambda r}{r^{2}}+\ldots
\end{eqnarray}
where $A^{(0)}$, $A^{(2)}$ and $\Lambda$ are integration constants.
The logarithmic term suffers from an ambiguity scale $\Lambda$,
which leads to a divergence in the Green's function.
But as explained in \cite{Hvarious}, this can be removed
by introducing an appropriate counter-term in the action.
From linear response theory, the conductivity $\sigma$
is related to the retarded current Green's function $G^{R}$ as follow,
\begin{eqnarray}
\label{conductivityD}
\sigma(\omega)=\frac{1}{i\omega}G^{R}(\omega,k=0),
\end{eqnarray}
where the retarded current Green's function $G^{R}$ can be
evaluated by using the standard AdS/CFT technique as\footnote{In Ref.\cite{JPWu},
the authors have given a detailed derivation of the retarded Green's function.
A similar calculation can be also found in Ref.\cite{Myers}.
They find that Weyl term has no explicit effect on the retarded Green's function and
its expression is the same as that in
Einstein theory for a standard Maxwell field.
Therefore, in the following calculation, we can safely use the Eq.(\ref{conductivityD}).}
\begin{eqnarray}
\label{conductivityD}
G^{R}(\omega)=-\lim_{r \rightarrow \infty} r^{3}f(r)\frac{\partial_{r}A_{x}(r,\omega)}{A_{x}(r,\omega)}.
\end{eqnarray}

Substituting the solution (\ref{Axinfty}) into the above equation,
one can easily obtain the explicit expression of the retarded Green's function
\begin{eqnarray}
\label{GreenFAdS5v1}
G^{R}(\omega)=2\frac{A^{(2)}}{A^{(0)}}-\frac{\omega^{2}}{2}.
\end{eqnarray}

Therefore, the conductivity is given by
\begin{eqnarray}
\label{conductivityE}
\sigma(\omega)=-\frac{i A^{(2)}}{\omega A^{(0)}}+\frac{i\omega}{2}.
\end{eqnarray}

In order to read off $A^{(0)}$ and $A^{(2)}$, we must solve Eq.(\ref{GMaxwellxEOM}),
Eqs.(\ref{ScalarEOM1}) and (\ref{GMaxwellEOM1})
numerically with the boundary condition (\ref{ingoingC}).
For definiteness, we set $c_{4}=0$ and explore the effect of
$\alpha$, $c_{\alpha}$ and $\gamma$ on the conductivity.
The numerical results of the frequency dependent conductivity
are illustrated in FIG.\ref{conductivity1} and \ref{conductivity2}.
The main characteristics are summarized as follows:

\begin{enumerate}

\item The imaginary part has a pole at $\omega=0$. According to the Kramers-Kronig relation,
this indicates that the real part contains
a delta function. In addition, from FIG.\ref{conductivity1}, it is easy to find that there exists a gap in the conductivity, which rises quickly near the gap frequency $\omega_{g}$.
They are two main common features of the standard version holographic superconductors\cite{3H,Hvarious}.

\item The ratio of gap frequency over critical temperature $\omega_{g}/T_{c}$ is unstable
and running with the model parameter $\alpha$, $c_{\alpha}$ and $\gamma$ (FIG.\ref{conductivity1}).
For fixed values of $\alpha$ and $c_{\alpha}$, the ratio $\omega_{g}/T_{c}$ increases
with the fall of the Weyl parameter $\gamma$. Especially, for $\gamma=0.04$,
the value of ratio $\omega_{g}/T_{c}\approx 6$, far less than $8$
found in the standard version holographic superconductors\cite{3H,Hvarious},
which is sharply different from the previous results. Therefore, compared with the former models, the value of our system is closer to the weakly coupled BCS value of $3.5$.
Moreover, if we fix values of Weyl parameter $\gamma$, the ration $\omega_{g}/T_{c}$ will increase
with the augmentation of model parameter $\alpha$ or $c_{\alpha}$.

\item In Ref.\cite{JPWu}, they have found that an extra spike
 appears inside the gap when $\gamma< 0$. But for this model, the new things are not only when $\gamma<0$ but also when $\alpha$ or $c_{\alpha}$ are large, the spike will emerge, too.
 Moreover, FIG.\ref{conductivity1} implies that under the joint effects of $\gamma$, $\alpha$ and $c_{\alpha}$, that is, when $\gamma<0$ and $\alpha$ and $c_{\alpha}$ are large, there will be more spikes and each of them appears to be sharper and narrower than those under the effect of one single model parameter.

\item The parameter $\gamma$, $\alpha$ and $c_{\alpha}$ together control the strength of fluctuations
in this system. For fixed value of $\gamma$, the coherence peak gradually becomes stronger and narrower with the increase of $\alpha$ or $c_{\alpha}$ (FIG.\ref{conductivity2}),
which is in agreement with the result found in Refs.\cite{stuckelberg2,stuckelberg4}.
For fixed value of $\alpha$ and $c_{\alpha}$, the coherence peak becomes stronger and narrower
with the fall of the Weyl parameter $\gamma$ and the fluctuations
will be suppressed completely at larger value of $\gamma=0.04$ (FIG.\ref{conductivity2}).

\end{enumerate}

\section{Conclusions and discussion}

In this paper, we have constructed the St\"{u}ckelberg holographic superconductor
with the Weyl corrections and obtained the numerical solutions of this model.
We examine carefully the effects of model parameters, $c_{\alpha}$, $\alpha$, $c_{4}$
and the Weyl coupling parameter $\gamma$ on the phase transitions.
For the second order phase transition, the critical temperature $T_{c}$ only depends
on the Weyl parameter and the form of $\mathcal{F}(\Psi)$ is not the determinant of $T_{c}$.
For fixed model parameter $c_{\alpha}$, $\alpha$ and $c_{4}$,
the critical temperature will be increasingly high as we amplify the parameter $\gamma$.
Therefore, the condensation gets harder when $\gamma< 0$.
But for $\gamma> 0$, the tendency is just opposite. It is
in line with the results found in Ref. \cite{JPWu}.

We specially focus on how the Weyl coupling parameter $\gamma$,
together with the model parameters $c_{\alpha}$, $\alpha$, $c_{4}$,
exerts an effect on the order of phase transitions
and the critical exponents of second order phase transitions.
We find that the Weyl coupling parameter $\gamma$ also changes the order of the phase
transition besides the model parameters $c_{\alpha}$, $\alpha$, $c_{4}$.
Furthermore, we conclude that for some definite form of $\mathcal{F}(\Psi)$,
the critical exponents $\beta$ is only dependent on the model parameter
but independent of the Weyl coupling parameter $\gamma$ in second order phase transitions.

Finally, we calculate the conductivity of the St\"{u}ckelberg holographic superconductor
with the Weyl corrections numerically and obtain that
for fixed model parameters $c_{\alpha}$, $\alpha$ and $c_{4}$,
the ratio $\omega_g/T_c$ is unstable and
becomes smaller when the Weyl coupling parameter $\gamma$ increases,
and even falls to approximately $6$,
which is far less than the standard value $8$.
It is in agreement with the discovery in Ref.\cite{JPWu}.
In addition, we also examine how the parameter $\gamma$, $\alpha$ and $c_{\alpha}$
control the strength of fluctuations in this model together.
We observed that the strength of fluctuations was adjusted not only by
the parameter $\alpha$ and $c_{\alpha}$ as found in Refs.\cite{stuckelberg1,stuckelberg4},
but also by the Weyl coupling parameter $\gamma$.

On the whole, the higher curvature interactions on the gravity side
broaden the class of the dual field theories one can holographically study.
Various holographic superconductors model taken into account the higher curvature interactions
are working intensively \cite{1,2,3,4,5,6,7,8}.
It may be interesting to further explore
the joint effects of various higher curvature interactions, including Gauss-Bonnet, quasi-topological and Weyl terms, on the holographic superconductors.

\begin{acknowledgments}

D. Z. Ma is supported by the Team Research Program of Hubei University for Nationalities(No. MY2010T001).
Y. Cao and J. P. Wu are partly supported by NSFC(No.10975017) and the Fundamental Research
Funds for the central Universities.

\end{acknowledgments}

\end{document}